\documentclass[%
 reprint,
 amsmath,amssymb,
 aps, nofootinbib
]{revtex4-2}

\usepackage{graphicx}
\usepackage{dcolumn}
\usepackage{bm}

\begin{document}

\preprint{}

\title{Code properties of the holographic Sierpinski triangle}

\author{Ning Bao}
\email{ningbao@bnl.gov}
\affiliation{Computational Science Initiative, Brookhaven
National Laboratory, Upton, New York 11973 USA
}

\author{Joydeep Naskar}
\email{naskar.j@northeastern.edu}
\affiliation{Department of Physics, Northeastern University, Boston, Massachusetts 02115, USA}

\date{\today}

\begin{abstract}
We study the holographic quantum error correcting code properties of a Sierpinski triangle-shaped boundary subregion in $AdS_4/CFT_3$. Due to existing no-go theorems in topological quantum error correction regarding fractal noise, this gives holographic codes a specific advantage over topological codes. We then further argue that a boundary subregion in the shape of the Sierpinski gasket in $AdS_5/CFT_4$ does not possess these holographic quantum error correction properties.
\end{abstract}

\maketitle


\section{\label{sec:level1}Introduction}

The exploration of gravity and quantum field theories from an information theory perspective has had a long and fruitful history. The study of holographic entanglement entropy \cite{rt-formula} eventually led to the study of holographic quantum error correction, first defined in \cite{harlow-first}. This development led to a series of influential work in AdS/CFT, in particular \cite{pastwaski-harlow}\cite{dong-entanglement-wedge}\cite{harlow-tasi}.

Quantum information science has been one of the most active areas of research in the last decades. A plethora of literature has been produced in pursuit of the physical realization of quantum computers. The necessity for robust information encoding that can withstand errors makes the development of quantum error correction very important. In this field, tremendous progress has been made in both the development of efficient quantum error correcting codes (QECC) and the creation of appropriate hardware candidates for their physical realisation.\cite{morten2019}

The area of topological phases of matter\cite{witten-topo-phase} has also been an area of active research over the past few decades. The combination of developments in that area and that in quantum error correction has led to the development of topological quantum error correcting codes\cite{bombin2013}. While such codes are quite efficient and powerful, a recent no-go theorem has revealed limitations of topological quantum error correction with regard to fractal noise \cite{arpit2021}. Fractal noise is a quite reasonable model for real-world experimental noise, due to defects on the lattice and percolation; see \cite{auger2017}\cite{percolation2017} for a detailed discussion. One of the main theorems of \cite{arpit2021} states:
\\
\textbf{Theorem:} \textit{$Z_N$ topological order cannot survive on a fractal
embedded in a 2D Euclidean space $\mathbb{R}^2$.}

It is therefore a natural question to ask if holographic quantum error correction suffers from the same limitation against fractal noise. In this paper, we answer this question in the negative. In this article, we discuss extension of so-called uberholography \cite{pastawski-preskill2016}, a prescient study of robustness of holographic QECC to fractal erasure noise in $AdS_3/CFT_2$ to $AdS_4/CFT_3$, in particular considering the quantum error correction properties of a boundary subregion in the shape of a Sierpinski triangle.

In particular, by taking a time slice in (2+1)d CFT, we essentially have a $\mathbf{R}^2$ surface, and if one is able to demonstrate bulk reconstructability of operators deep within the bulk, then this would show that holographic QECCs do not obey an analog to the topological QECC no-go theorem. 

The organization of the paper is as follows: In Sec. II, we give a brief background of holographic QECC, in Sec. III we will discuss the code properties of the Sierpinski triangle subregion, and we will conclude with some discussion and potential future work in Sec. IV.

\section{\label{sec:level2}Brief Background}
The $AdS_{d+1}/CFT_d$ correspondence is the duality between a theory of quantum gravity in d+1 space-time dimensions and that of a conformal field theory living on its boundary. Primaries in the CFT can for example be mapped to bulk fields, using an extrapolate dictionary, and many other entries of this holographic dictionary exist. For a review of the AdS/CFT correspondence, refer to \cite{maldacena-adscft}\cite{maldacena-adscft-extensive}\cite{polchinski-adscft}.

\subsection{Minimal surfaces and Ryu-Takayanagi formula}
The Ryu-Takayanagi formula\cite{rt-formula} relates minimal surface area in the AdS bulk to entanglement entropy of geometric subregions of the boundary CFT at leading order is given by
\begin{equation}
    S_A= \frac{|\chi_A|}{4G_N},
\end{equation}
where $\chi_A$ is the minimal surface in the bulk the of the curve $\partial A$ homologous to A on the boundary and $|\chi_A|$ is its area.

The Ryu-Takayanagi formula requires corrections when the boundary subregion has sharp corners. It was found in \cite{tonni-2014} that minimal surfaces in $AdS_4$ with finite number of vertices in their corresponding boundary subregions have area

\begin{equation}
    |\chi_A|= \frac{P_A}{a}-B_A \log{\frac{P_A}{a}}-W_A +o(1),
    \label{eq:AdS4-minimal-area}
\end{equation}
where $a$ is a length-scale cutoff and $P_A$ is the perimeter of the curve $\partial A$. Both $W_A$ and $B_A$ are parameters of the shape of the region as explained in \cite{tonni-2014}. The term $W_A$ is subleading, and thus can be neglected here, while $B_A$ is the leading order term of $\Tilde{B_A}$, defined to be:
\begin{equation}
    \begin{split}
        & \Tilde{B_A}=\frac{1}{\log({a/P_A})}\left( |\chi_A|-\frac{P_A}{a}\right)\\
        & \Tilde{B_A}=B_A+o(1)
    \end{split}
\end{equation}
The $o(1)$ terms vanish as $a\rightarrow 0$.

\subsection{Review of holographic codes in $AdS_3/CFT_2$}
Here we will briefly review the relevant background on holographic QECC. After the development of the general theory of holographic QECC in \cite{harlow-first}, code properties of holographic geometries in $AdS_3/CFT_2$ were studied in \cite{pastawski-preskill2016}. It has been shown that bulk operators deep in the center of AdS space can be recovered, even when the support on the boundary region is given by a Cantor set of measure zero. This unexpected recoverability is known as uberholography. 

As mentioned before, the boundary subregion in this case is a Cantor set of disconnected points.
The fractal nature of the boundary subregion plays a crucial role in this construction, and we will review precisely how this works in this subsection.
\\

The minimal surface of a connected region $R$ in a 1d boundary slice is actually the bulk geodesic $\chi_R$, and its "area" is the length of the geodesic. For a boundary region $R$ with length $|R|$, the minimal area $|\chi_R|$ is given by
\begin{equation}
    |\chi_R|= 2L \log{\frac{|R|}{a}},
\end{equation}
where $L$ is the radius of curvature of the hyperbolic geometry and $a$ is the short-distance cutoff, that we have encountered previously in Eq. \ref{eq:AdS4-minimal-area}.

\begin{figure}[h!]
		\centering
		\includegraphics[width=0.5\textwidth]{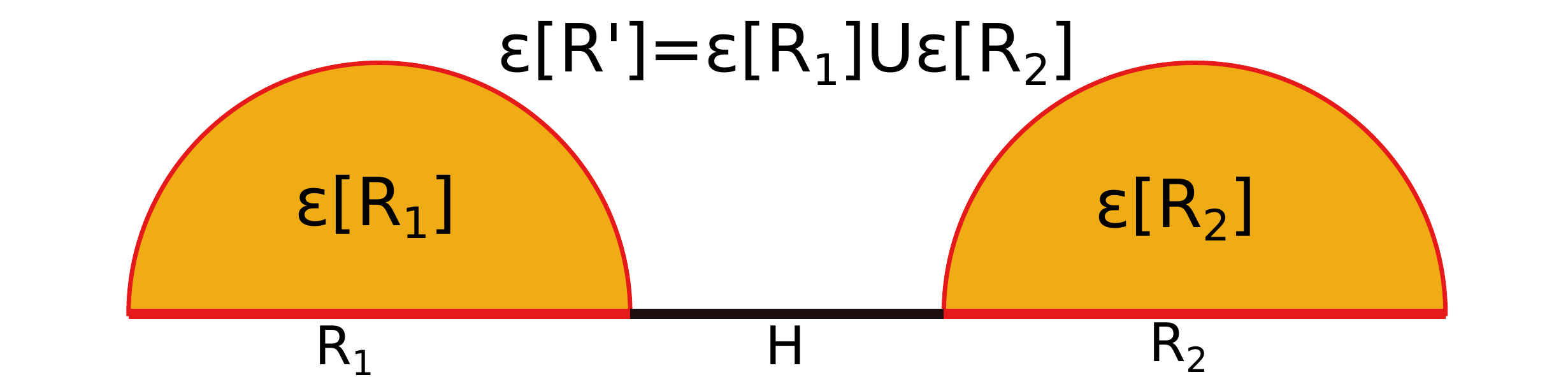}
	\caption{Surface $\chi_{R'}=\chi_{R1}\cup\chi_{R2}$. Disconnected regions $R_1$ and $R_2$ are drawn in red. The hole $H$ is drawn in black. The shaded region is the entanglement wedge $\epsilon[R']=\epsilon[R_1]\cup \epsilon[R_2]$.}
	\label{fig:1dfractcal-example1.1}
\end{figure}

\begin{figure}[h!]
		\centering
		\includegraphics[width=0.5\textwidth]{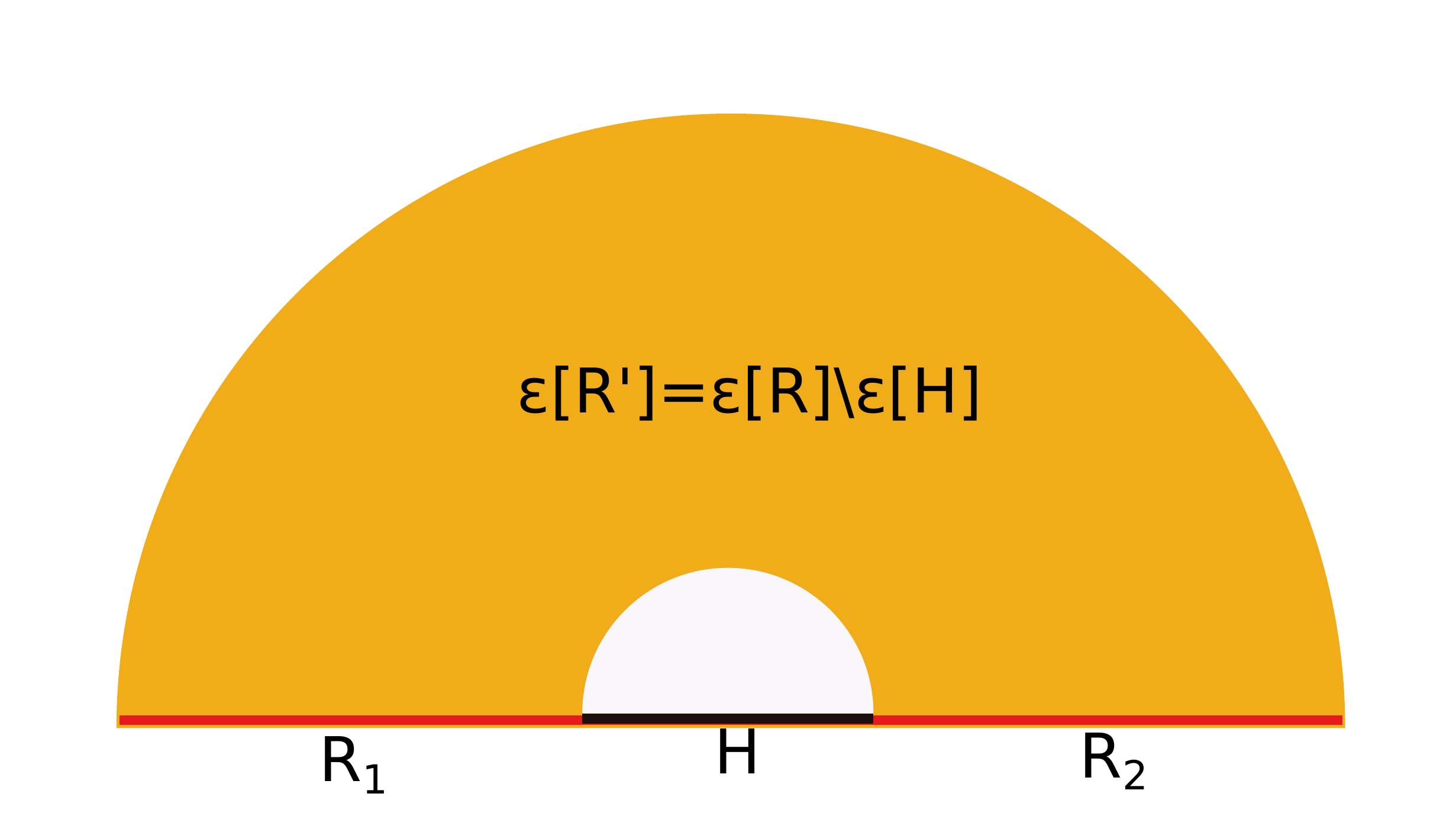}
	\caption{Surface $\chi_{R'}=\chi_{R}\cup\chi_{H}$. Disconnected regions $R_1$ and $R_2$ are drawn in red. The hole $H$ is drawn in black. The shaded region is the entanglement wedge $\epsilon[R']=\epsilon[R]\backslash \epsilon[H]$.}
	\label{fig:1dfractcal-example1.2}
\end{figure}

We begin by considering a boundary region $R$ with a hole $H$, such that they are divided into three parts: two disjoint boundary regions $R_1$, $R_2$ and the hole $H$ such that
\begin{equation}
    |R_1|=|R_2|=(\frac{r}{2})|R|, \quad |H|=(1-r)|R|.
\end{equation}
Now the boundary region $R'$ is disconnected
\begin{equation}
    R'=R_1\cup R_2 = R \backslash H.
\end{equation}
There are two ways to choose the bulk geodesics $\chi_R'= \chi_{R_1} \cup \chi_{R_2}$ or $\chi_R'=\chi_R\cup\chi_H$, with their respective entanglement wedges $\epsilon[R']=\epsilon[R_1]\cup \epsilon[R_2]$ and $\epsilon[R']=\epsilon[R] \backslash \epsilon[H]$, respectively (see Figs. \ref{fig:1dfractcal-example1.1},\ref{fig:1dfractcal-example1.2}).
We will be working in the regime where
\begin{equation}
    |\chi_{R_1}|+|\chi_{R_2}|>|\chi_{R}|+|\chi_{H}|.
\end{equation}
Saturating this bound gives us that $\frac{r}{2}=\sqrt{2}-1$. Each component of $R'$ is smaller than $R$ by $r/2$.
Now let us iterate making holes, until the size of each component is reduced to the cutoff length $a$. Let us say, we arrived at this configuration after $m$ steps. This gives
\begin{equation}
    a=\left(\frac{r}{2}\right)^m |R|.
\end{equation}
We call the remaining region $R_{min}$. It has $2^m$ components each of length a. We define the distance of the code with operator algebra $A$ in bulk region X to be 
\begin{equation}\label{code-dist-2d}
d(A_X)\leq \frac{|R_{min}|}{a}=2^m=\left(\frac{|R|}{a}\right)^{\alpha},
\end{equation}
where
\begin{equation}
    \alpha=\frac{\log{2}}{\log{2/r}}=\frac{1}{\log_2{(\sqrt{2}+1})}=0.786,
\end{equation}
so the distance is bounded above by some $n^{\alpha}$.

While uberholography gave a good characterization of holographic codes with fractal geometries in 2d CFTs, it remains an open question for study in the context of holographic 3d CFTs\footnote{While this article was under review, \cite{ageev-2022} appeared where the author studied uberholography in higher dimensions for Cantor-set like erasures.}. In particular, a further motivation for pursuing this question is to compare the performance of holographic and holographic-inspired codes over topological codes, particularly in the context of the fractal noise no-go theorem of \cite{arpit2021}, in particular that there cannot be any topological codes robust against fractal noise embedded on a flat 2d plane. The ability to construct holographic codes that do not possess this limitation is therefore of clear interest, particularly in the context of $AdS_4/CFT_3$.

\section{The Holographic Sierpinski Triangle}
\begin{figure}[h!]
		\centering
		\includegraphics[width=0.5\textwidth]{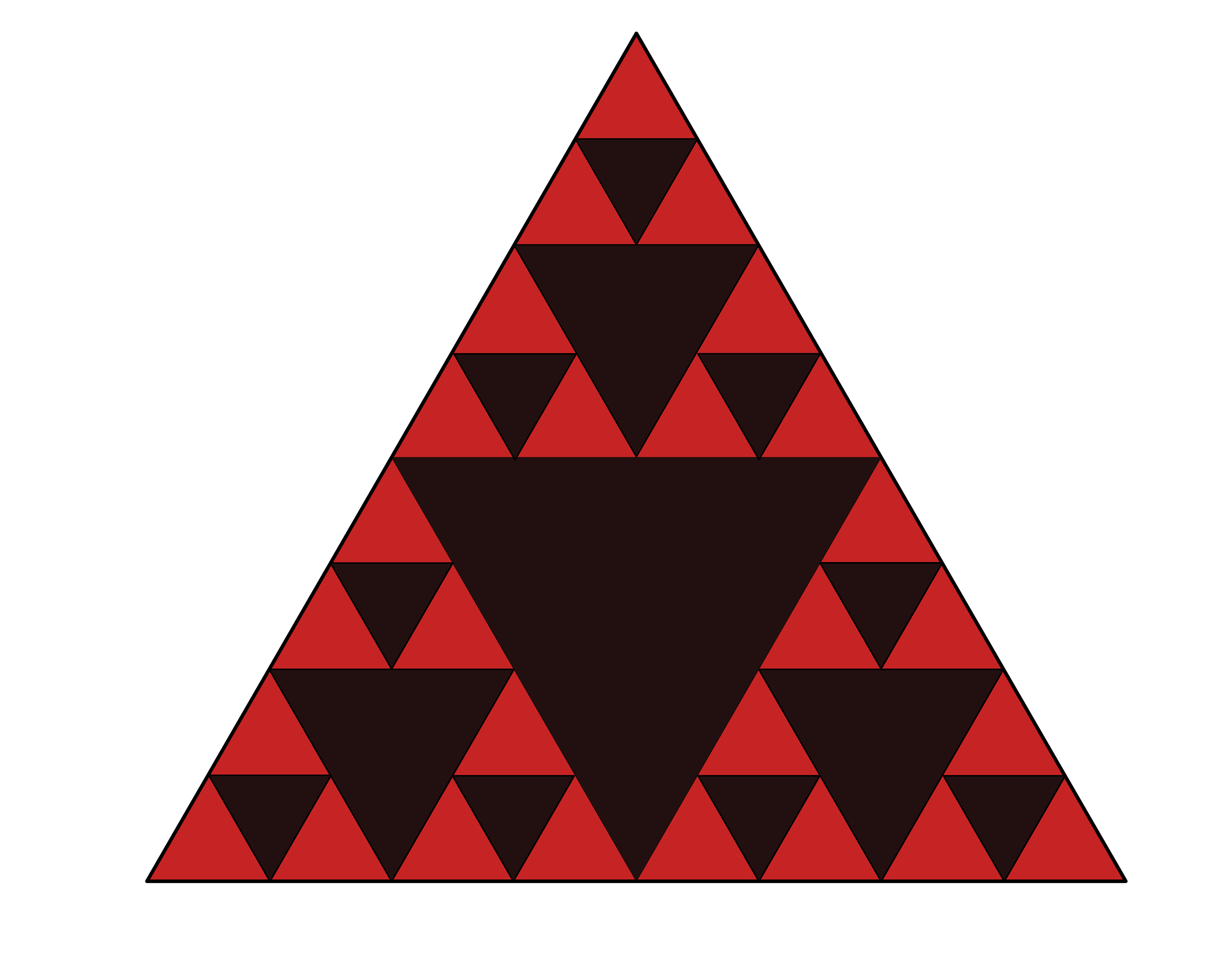}
	\caption{The Sierpinski Triangle. The fractal is constructed by removing triangular holes (shaded black) of decreasing size from the big triangle. The remaining area (shaded red) is the boundary subregion, whose measure goes to zero as $m\rightarrow\infty$.}
	\label{fig:sierpinski-triangle}
\end{figure}

The Sierpinski triangle is a fractal geometry with Hausdorff dimension 1.585, as shown in fig \ref{fig:sierpinski-triangle}.

\begin{figure}[h!]
		\centering
		\includegraphics[width=0.5\textwidth]{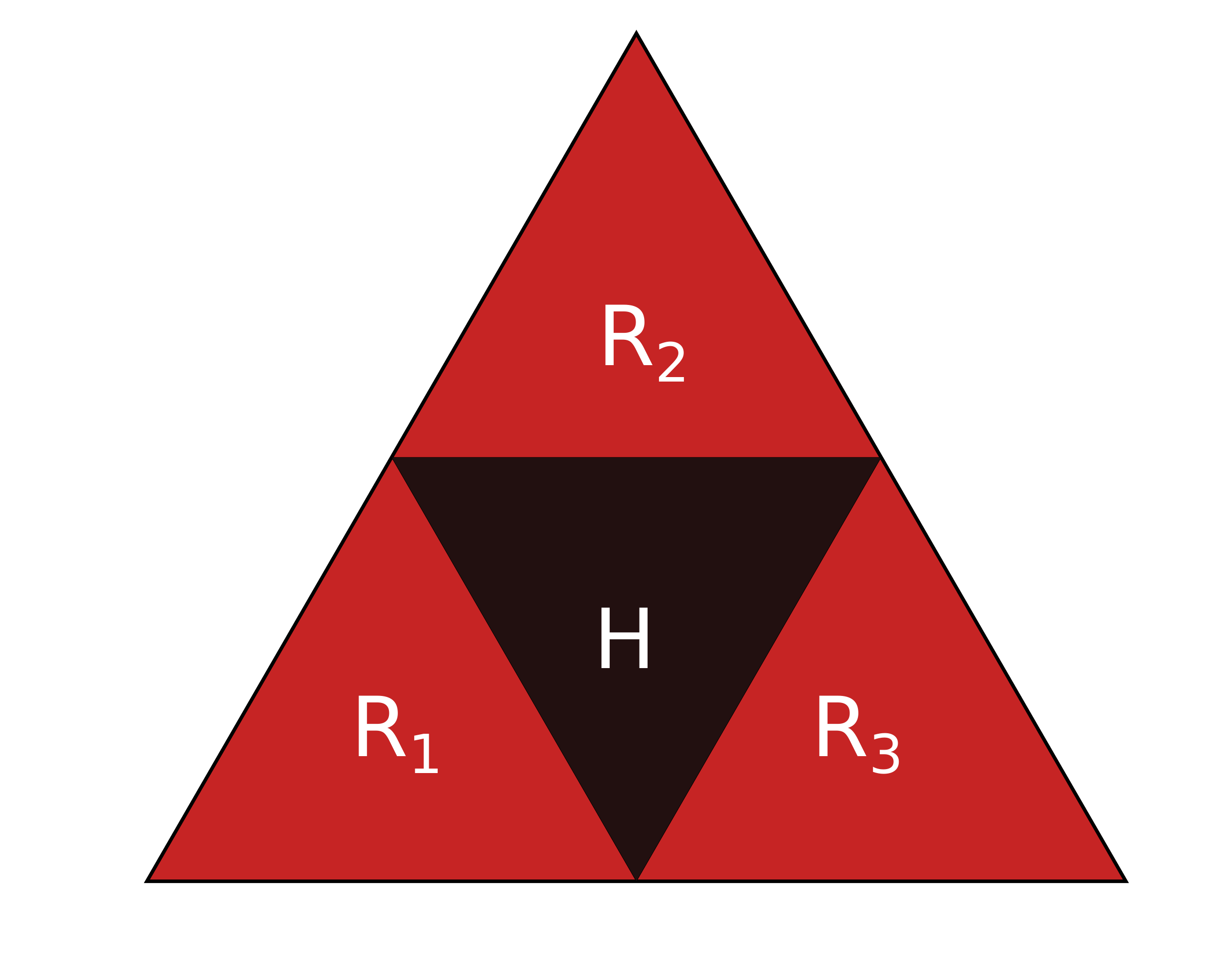}
	\caption{A triangular subregion $H$ removed from the big triangle R, leaving three triangular subregions $R_1$, $R_2$ and $R_3$ on the boundary. $R'=R_1\cup R_2 \cup R_3$ is the relevant boundary subregion of our interest. By geometric complementarity, $R=R' \cup H$. In the previous section, we denoted $R$(physical boundary) by $\Phi$ and $R'$(boundary subregion) by $R$.}
	\label{fig:triangle-step1}
\end{figure}

For the case of an $N$-sided regular polygon of side length $l$, the calculation of minimal area has some simplifications \cite{tonni-2014}
\begin{equation}
\begin{split}
& P_A= Nl, \\
& B_A= 2Nb(\alpha_N), \\
& \alpha_N= \frac{N-2}{N}\pi,
\end{split}    
\end{equation}
where $b(\alpha)$ is a regulator-independent coefficient that depends on the opening angle $\alpha$ as defined in \cite{tonni-2014}. See \cite{myers-2015} for universality and CFT interpretation of this factor.
Consider the the disconnected boundary region in figure \ref{fig:triangle-step1}, where a region $H$ given by an equilateral triangle of side $l_1$ (or area $\mathcal{A}_{l_1}$) has been carved out from the center of a bigger triangle of side $l_0$(or area $\mathcal{A}_{l_0}$), labeled $R$. Note that $l_1=\frac{l_0}{2}-\epsilon$ where $\epsilon>0$ is extremely small compared to $l_0$. We will eventually fix $\epsilon$ (to be of the order of distance cutoff $a$) to satisfy the condition for error correction.

Consider the boundary subregion $R'=R \backslash H$ and $R'=R_1 \cup R_2 \cup R_3$. They represent the same boundary subregion with different minimal surfaces in the bulk, namely $\chi(R')_{con.}=\chi(R) \cup \chi(H)$ and $\chi(R')_{disc.}=\chi(R_1) \cup \chi(R_2) \cup \chi(R_3)$, respectively. While one of these entanglement wedges is disconnected, the other is connected. The minimal surface area is the lesser of these two. 

Since $\epsilon$ is very small, we consider the disconnected wedge $\chi(R')_{disc.}=\chi(R_1) \cup \chi(R_2) \cup \chi(R_3)$ has a surface area roughly given by
\begin{equation}
    \begin{aligned}
       |\chi(R')|_{disc.}& = |\chi(R_1)|+|\chi(R_2)|+|\chi(R_3)| \\ &= 3\left[ \frac{3l_0}{2a}-6 b\left(\frac{\pi}{3}\right)\log{\left(\frac{3l_0}{2a}\right)}\right]\\
       &=\frac{9}{2} \frac{l_0}{a} - 6 b\left(\frac{\pi}{3}\right)\log{\left(\frac{27l_0^3}{8a^3}\right)}.
    \end{aligned}
\end{equation}
Note that in the above expression we have assumed that the evaluated area is the same as it would be in the case $\epsilon=0$ and it suffices for our analysis. The connected wedge $\chi(R')_{con.}=\chi(R) \cup \chi(H)$ has a surface with area:
\begin{equation}
\begin{split}
    |\chi(R')|_{con.} =& |\chi(R)|+|\chi(H)| \\
     =& \frac{3l_0}{a}+\frac{3(l_1-\epsilon)}{a}\\ 
    & -6 b\left(\frac{\pi}{3}\right) \left[\log{\left(\frac{3l_0}{a}\right)}+\log{\left(\frac{3(l_1-\epsilon)}{a}\right)}\right] \\
     =& \frac{9}{2} \frac{l_0}{a} - 6 b\left(\frac{\pi}{3}\right) \log{\left(\frac{9l_0^2}{2a^2}\right)}\\
     & -\frac{3\epsilon}{a}-6 b\left(\frac{\pi}{3}\right)\log{\left(1-\frac{2\epsilon}{l_0}\right)}\\
     =& \frac{9}{2} \frac{l_0}{a} - 6 b\left(\frac{\pi}{3}\right) \log{\left(\frac{9l_0^2}{2a^2}\right)}\\
     & -\frac{3\epsilon}{a}+6b\left(\frac{\pi}{3}\right)\frac{2\epsilon}{l_0}
\end{split}
\end{equation}
where in the last equality we have used the series expansion $\log{(1+x)}\sim x$ for $x<<1$.
The condition for reconstructability is that the connected wedge should have the minimal area
\begin{equation}\label{qecc-condition}
    |\chi(R)|+|\chi(H)|\leq |\chi(R_1)|+|\chi(R_2)|+|\chi(R_3)|
\end{equation}
Neglecting the $\epsilon/l_0$ term, this inequality is satisfied when 
\begin{equation}
    \epsilon\geq 2ab\left(\frac{\pi}{3}\right)\log{\frac{3l_0}{4a}}
\end{equation}
We notice that at leading order, the areas are equal when $\epsilon=0$ (This particular choice of ratios of characteristic sizes of $R$ and $H$ is to leading order a phase transition in the entanglement wedges of the Sierpinski triangle). The comparison thus comes down to sub-leading order. We can think of $\epsilon$ as a regulator to preserve error correction properties.

Consider that after $m$ such iterations, the smallest triangle has a side of length scale $a$. We have removed triangles of various side length. The smallest triangle has:
\begin{equation}
\begin{split}
&l_m=\left(\frac{1}{2}\right)^m l_0 =a,
\end{split}    
\end{equation}
where $\mathcal{A}_a$ is the area of the smallest triangle in the boundary.

After $m$ steps, we have the disconnected wedge has an area
\begin{equation}\label{discon-m-iters}
    |\chi(R')|_{disc.}=3^m \left(\frac{3l_0}{2^m a}\right)-3^m 6b\left(\frac{\pi}{3}\right)\log{\left(\frac{3l_0}{2^m a}\right)}
\end{equation}
while the area of the connected wedge is given by
\begin{equation}\label{conn-m-iters}
\begin{split}
|\chi(R')|_{con.} = & |\chi(R)|+|\chi(H_1)|+ 3|\chi(H_2)|+ 3^2 |\chi(H_3)|+ \\ & \dots + 3^{m-1} |\chi(H_m)|  \\
 = & |\chi(R)|+ \sum_{j=1}^{m}3^{j-1}|\chi(H_j)| \\
 = & \left[\frac{3l_0}{a}+ \sum_{j=1}^m 3^{j-1}\frac{3l_0}{2^j a}\right] -\frac{3\epsilon}{a}\sum_{j=1}^m 3^{j-1}\\ &
-6 b\left(\frac{\pi}{3}\right) \left[\log{\left(\frac{3l_0}{a}\right)}+ \sum_{j=1}^m 3^{j-1}\log\left({\frac{3l_0}{2^j a}}\right)\right]\\
& -6 b\left(\frac{\pi}{3}\right)\left[\sum_{j=1}^m 3^{j-1}\log\left(1-\frac{2^j\epsilon}{l_0}\right) \right]
\end{split}
\end{equation}
The construction has to satisfy the level-$m$ analog of (\ref{qecc-condition}). We notice that the first term of (\ref{discon-m-iters}) equals the first term of (\ref{conn-m-iters}) as the latter is a finite GP series. The difference between second term of (\ref{discon-m-iters}) and third term of (\ref{conn-m-iters}) is equal to
\begin{equation}
    -6b\left(\frac{\pi}{3}\right) \log\left[\frac{1}{2^{\frac{m}{2}(m+1)}} \left( \frac{3\left(\frac{l_0}{2^m}\right)}{a} \right)^{\frac{1}{2}(3^m - 1)} \right]
    \label{difference-term-m-iter}
\end{equation}
As we are in the regime $l_0/2^m>>\epsilon$, we can neglect the last term of (\ref{conn-m-iters}) in comparison to its second term. This leaves us with
\begin{equation}
    \begin{aligned}
    \epsilon \geq 2ab\left(\frac{\pi}{3}\right)\left[\log{\left(\frac{3l_0}{2^m a}\right)}-\frac{m(m+1)}{3^m}\log{2}\right]
    \end{aligned}
\end{equation}
Dropping the second term, the critical value of $\epsilon$ for the connected phase to dominate is
\begin{equation}
    \epsilon=2ab\left(\frac{\pi}{3}\right)\log{\left(\frac{3l_0}{2^m a}\right)}
\end{equation}

As the limiting case\footnote{In this limiting case, $\epsilon=a\left(2b(\pi/3)\log{3}\right)$.}, on the boundary, after $m$ iterations, there are $3^m$ triangles of side $a=(\frac{1}{2})^m l_0$ remaining. Each such triangle has an area $\mathcal{A}_a = \frac{\sqrt{3}}{4}a^2.$ So the remaining area is
\begin{equation}
    \mathcal{A}_{min}=3^m \mathcal{A}_a = 3^m \frac{\sqrt{3}a^2}{4} = \frac{\sqrt{3}l_0^2}{4}\frac{3^m}{4^m}.
\end{equation}
The code \emph{distance} in the context of \cite{pastawski-preskill2016} is defined in equation (\ref{code-dist-2d}). Analogously in our case, the definition of \emph{distance} of the code with operator $A$ in bulk region $X$ is
\begin{equation}
    d(A_X) \leq \frac{\mathcal{A}_{min}}{\mathcal{A}_a} = 3^m = \left(\frac{\mathcal{A}_{l_0}}{\mathcal{A}_a}\right)^{\alpha},
\end{equation}
where
\begin{equation}
    \alpha=\frac{\log{3}}{\log{4}}=\frac{1}{\log_3{4}}=0.7925.
\end{equation}

The value of $\alpha$ is exactly half of that of the Hausdorff dimension. We believe that this factor 1/2 is related to the dimensionality of the embedding space, which in this case is 2.
The fact that the connected phase is the minimal surface for the Sierpinski triangle boundary region guarantees that an operator located deep in the bulk is reconstructible even if the boundary region one has access to is that of the Sierpinski triangle. This is an immediate example of an instance in which the holographic quantum error correcting code can handle fractal erasure noise, in a way that topological QECC were proven unable to do so in \cite{arpit2021}.

\subsubsection*{Sierpinski gasket as a boundary 4d $CFT$}
The code properties of the Sierpinski triangle-shaped boundary subregion is, however, limited to 3d $CFT$\footnote{Note that here we ignore the trivial extension of tensoring the Sierpinski triangle to $\mathbb{R}_1$, to promote them to "strips;" this would certainly work, but is not particularly natural in fractal sense.} (and, if you like, 2d $CFT$ via uberholography). The leading term in the expression for minimal area was a linear one, which turned out to be equal for both candidates for the minimal surface, paving the way for comparison at sub-leading order which favoured the surface $\chi(R')_{con.}=\chi(R)\cup\chi(H)$ as the minimal surface over $\chi(R')_{disc.}=\chi(R_1)\cup\chi(R_2)\cup\chi(R_3)$.
Recall that the RT formula gives that the leading term of the minimal area in the (d+1)-dimensional bulk enclosed by a d-dimensional boundary CFT region scales as the order of co-dimension 2 of the time-slice of (d+1)-dimensional bulk.
If we consider the Sierpinski Gasket as a 4d boundary $CFT$, time-slices of the bulk are now 4-dimensional hyperplanes and co-dimension 2 "surfaces" of the same no longer scale linearly. The leading order terms therefore favours the disconnected phase, and the ability to reconstruct operators deep in the bulk interior ceases.
However, this does not limit the code properties of other fractal geometries in higher dimensions such as Cantor-like slicing in a special direction or orientation as in \cite{ageev-2022}.

\section{Conclusion}
To summarise our work, we have studied the holographic QECC properties of a boundary region in the shape of a Sierpinski triangle in $AdS_4/CFT_3$, where our boundary region was precisely a fractal embedded in a flat 2 dimensional plane. This is relevant for mainstream quantum computation because, while topological codes are the current state of the art, holography-inspired codes seem to have at least one advantage over them, specifically that holographic codes can be robust against fractal noise while topological codes cannot in three dimensions. That said, we are working in the large N limit, which is infeasible in real life; it is possible that subleading corrections would change this story. We have also argued that this propertie does not generalize to $AdS_5/CFT_4$. However, topological codes with fractal geometries are more easily constructed in higher dimensions, whereas holographic codes dominate in lower dimensions. This presents an elegant conceptual picture for when one may favor holographic or holographic-inspired QECC's over their topological cousins.

There are a few potential directions for future study. First, one could study boundary subregions of other fractal shapes, and ask if they also have nice bulk reconstruction properties. Secondly, one can study whether other fractals in higher dimensions have nice bulk reconstruction properties, if they are generalized from lower dimensional analogs that are not Sierpinski. Finally, one could build a practical holographic-inspired QECC and run it on near-term quantum hardware to experimentally demonstrate these robustness properties.

\begin{acknowledgments}
J.N. would like to thank Swati Chaudhary for introducing to graphics packages Affinity and Inkscape. J.N. and N.B. would like to thank Chris Akers and Guanyu Zhu for useful comments on the draft.  N.B. was supported by the Department of Energy under grant number DE-SC0019380, and is supported by the Computational Science Initiative at Brookhaven National Laboratory, and by the U.S. Department of Energy QuantISED Quantum Telescope award. J.N. is supported by the Graduate Assistantship from the Department of Physics, Northeastern University.
\end{acknowledgments}

\nocite{*}


\end{document}